\documentclass{aa}

\def\degr{$^{\circ}$}

\usepackage{natbib}
\bibpunct{(}{)}{;}{a}{}{,}

\usepackage{times}
\begin{document}
\title{The Hipparcos observations and the mass of sub-stellar objects\thanks{Based on observations from the Hipparcos astrometric satellite operated by the European Space Agency (ESA 1997)}}
\titlerunning{Hipparcos and sub-stellar objects}
\author{D.~Pourbaix\thanks{Postdoctoral Researcher, F.N.R.S., Belgium}\inst{1,2}}
\institute{Institut d'Astronomie et d'Astrophysique, 
Universit\'e Libre de Bruxelles, C.P.~226, Boulevard du Triomphe, B-1050 
Bruxelles, Belgium
\and
Department of Astrophysical Sciences, Princeton University, Princeton, NJ 08544-1001, U.S.A.}
\date{Received date; accepted date} 
\offprints{D.~Pourbaix \email{pourbaix@astro.ulb.ac.be}}
\abstract{
The Hipparcos Intermediate Astrometric Data have been used lately to estimate the inclination of the orbital plane of candidate extrasolar planets.  Whereas most of these investigations derive almost face-on orbits, we show that the astrometric data are seldom precise enough to undertake such studies and that the `face-on' result might be just a spurious effect of the method.
}
\maketitle

\section{Introduction}
Even before the release of the Hipparcos and Tycho Catalogues \citep{Hipparcos}, their interest regarding the extrasolar planets was already pointed out \citep{Perryman-1996:a}.  For the past two years, some investigators have re-processed the Hipparcos Intermediate Astrometric Data (IAD) of different classes of objects.  Though initially limited to binaries \citep{Soderhjelm-1999,Halbwachs-2000:a,Pourbaix-2000:b}, the planet hunters have also got interested in such re-processings \citep{Mazeh-1999:a,Gatewood-2001:a,Zucker-2000:a}.  A recent statistical study based on the IAD concluded that a significant fraction of the orbits of extrasolar planets are seen almost face-on \citep{Han-2001:a}, thus pushing up their mass estimates.  According to that study, a substantial number of these objects would no longer be planets.

We have re-analyzed the IAD of 46 stars with planetary candidates and found that the orbital model, with the spectroscopic parameters assumed, seldom improves the astrometric fit significantly.  We show that in most cases the small inclinations found by \citet{Han-2001:a} are just an artifact of the fitting procedure.  Such a caveat is indeed related to the size of the orbit with respect to the precision of the astrometric measurements.  Similar results, although wrong, would therefore be derived with classes of instruments of the same given precision (e.g. MAP \citep{Gatewood-1987:a}).    

\section{Astrometric model and the choice of the parameters}\label{Sect:model}

Regardless of its physical nature, the motion of any member of a binary system with respect to the common center of mass is described by the well-known relations \citep{PrDoSt}
\begin{eqnarray}
x&=&A(\cos E-e)+F\sqrt{1-e^2}\sin E,\\
y&=&B(\cos E-e)+G\sqrt{1-e^2}\sin E
\end{eqnarray}
where $E$ is the eccentric anomaly, solution of Kepler's equation for given $e$, $P$, and $T$.  Seven orbital parameters are thus required: $A$, $B$, $F$, $G$, $e$, $P$, and $T$.  The first four, the Thiele-Innes constants, are linked to the Campbell elements ($a_a$, $i$, $\omega_1$, and $\Omega$) through
\begin{eqnarray}
A&=&a_a(\cos\omega_1\cos\Omega-\sin\omega_1\sin\Omega\cos i),\\
B&=&a_a(\cos\omega_1\sin\Omega+\sin\omega_1\cos\Omega\cos i),\\
F&=&a_a(-\sin\omega_1\cos\Omega-\cos\omega_1\sin\Omega\cos i),\\
G&=&a_a(-\sin\omega_1\sin\Omega+\cos\omega_1\cos\Omega\cos i).
\end{eqnarray}
There is {\em a priori} no mathematical reason to prefer the Campbell elements over the Thiele-Innes ones.  However, the spectroscopic orbit yields, among other quantities, $\omega_1$ and the product $a_l\sin i$ ($a_l$ is here reckoned in linear units whereas $a_a$ is its angular counterpart).  Therefore, it is convenient to use Campbell's elements even if global optimization techniques are then required to minimize $\chi^2$ \citep{Pourbaix-2000:b} because the model is highly non-linear in terms of the remaining elements.

Most of the sub-stellar objects have been detected through radial velocity surveys and, therefore, the spectroscopic orbit of the primary is known.  Coupled to the parallax ($\varpi$), two parameters thus remain unknown, namely $i$ and $\Omega$.  Indeed, $a_a$ can be written as
\begin{equation}
a_a=3.35729138\,10^{-5}K_1 P\sqrt{1-e^2}\varpi/\sin i,\label{Eq:defaa}
\end{equation}
where $K_1$ is the amplitude of the radial velocity curve reckoned in m~s$^{-1}$, $P$ is the period in yr, $\varpi$ in mas and $e$ is the eccentricity.

The IAD were made available to allow some further processing of the Hipparcos observations, especially those of binaries.  Regardless of the model used to derived the final Hipparcos results, the IAD are residuals with respect to the single star model.  When the binary nature of a star was discovered after the release of the Catalogues, the IAD can therefore be re-processed with an orbital model \citep{Halbwachs-2000:a,Pourbaix-2000:b}.  Although they are applied to 1-dimensional data, the equations of the orbital perturbation are essentially those of the visual absolute orbit.  They thus supply with the inclination $i$.

It is very tempting to re-process the IAD of all the stars around which planets were lately discovered.  Indeed, the motion of these stars should reveal tiny astrometric perturbations which can now be explained by the planet orbiting them.  Fitting the IAD should supply with the inclination and therefore with the mass of the companion.  For instance, that technique was applied to HD~10697 (HIP~8159) to obtain the mass of a brown dwarf \citep{Zucker-2000:a}.  The brown dwarf nature of the companion was derived from the rather low value of the inclination, thus leading to a quite large mass for the secondary.

\section{Inclination from a tiny orbital wobble}\label{Sect:tinywobble}

Why does fitting $i$ and $\Omega$ yield almost face-on orbits whereas, from a naive reasoning, one would expect preferentially edge-on situations for spectroscopically detected companions?  \citet{Halbwachs-2000:a} barely mention that feature.  Indeed, in order to validate their approach, they process the IAD of known single stars with the orbital model (with $\omega_1$, $e$, $P$, and $T$ fixed) and investigate the behavior of $a_a$.  Instead of $a_a=0$ (the real value for single stars), they derive its Rayleigh distribution and conclude that $a_a$ is slightly overestimated (when the same method is applied to genuine binaries).

They also noticed that the estimated $a_a$ is directly related to the residuals of the coordinates of the star, i.e. about 1 mas for Hipparcos.  Fitting an orbit to the IAD when $a_a\sin i \ll 1$~mas always leads to small inclinations: the smaller the product, the smaller the inclination.  This is only due to the relative precision of the instrument with respect to the magnitude of the orbital wobble.  The criterion depends on $a_a\sin i$, not on $K_1$.  Indeed, long-period astrometric binaries also characterized by small $K_1$ would be detected with a much longer Hipparcos-type mission because, {\em mutatis mutandis}, $a_a\sin i$ would increase with $P$.  The orbital periods of known sub-stellar candidates are simply too small for $a_a\sin i$ to be large enough.

What about the `precision' of such inclinations?  Let us have a look at the derivative of the Thiele-Innes constant $A$ with respect to $i$ (the reasoning is the same for $B$, $F$, and $G$).  These derivatives are indeed used to build the Fisher matrix whose inverse is the covariance matrix of the fitted parameters \citep{DaReErAnPhSc}.
\begin{eqnarray}
\frac{\partial A}{\partial i}&=&\frac{\partial a_a}{\partial i}(\cos\omega_1\cos\Omega-\sin\omega_1\sin\Omega\cos i) \nonumber \\ 
&&{}+a_a\sin\omega_1\sin\Omega\sin i.
\end{eqnarray}
If $a_a$ is an independent parameter in the astrometric solution, the first term disappears and the second vanishes with $i$.  In that case, the smaller the inclination, the larger its uncertainty.  On the other hand, when Eq.~(\ref{Eq:defaa}) is adopted as a constraint,
\begin{equation}
\frac{\partial a_a}{\partial i}= -a_a\cot i.
\end{equation}
Hence, the smaller the inclination, the larger the derivatives, the smaller the standard deviation of the inclination, regardless of the precision of the astrometric data.

\section{Do these orbits improve the fit?}\label{Sect:improvement}

\begin{table*}[htb]
\caption[]{\label{Tab:Planets}List of planet candidates and their $a_a\sin i$ based on the spectroscopic orbit and the Hipparcos parallaxes. $i$ is the inclination derived from the IAD when $a_a\sin i$ is used as a constraint.  $\alpha$ is the probability of obtaining an $F$-value greater or equal to $\hat{F}$ if there is no orbital wobble present in the IAD.  Ref is the reference for the orbit: 1: \citet{Naef-2000:a}; 2: \citet{Queloz-2000:a}; 3: \citet{Butler-1999:a}; 4: \citet{Vogt-2000:a}; 5: \citet{Fischer-2001:a}; 6: \citet{Queloz-2000:b}; 7: \citet{Marcy-2000:a}; 8: \citet{Kurster-2000:a}; 9: \citet{Hatzes-2000:a}; 10: \citet{Butler-2000:a}; 11: \citet{Butler-1997:a}; 12: \citet{Naef-2001:a}; 13: \citet{Mayor-2000:a}; 14: \citet{Korzennik-2000:a}; 15: \citet{Butler-1996:a}; 16: \citet{Marcy-1999:a}; 17: \citet{Marcy-1996:a}; 18: \citet{Udry-2000:b}; 19: \citet{Noyes-1999:a}; 20: \citet{Udry-2000:a}; 21: \citet{Cochran-1997:a}; 22: \citet{Sivan-2000:a}; 23: \citet{Mazeh-2000:a}; 24: \citet{Delfosse-1998:a}; 25: \citet{Mayor-1995:a}}
\setlength{\tabcolsep}{2mm}
\begin{tabular}{llllcr|llllcr}\hline
HIP & HD/ & Ref & $a_a\sin i$ & $i$ & $\alpha$ & HIP & HD/ & Ref & $a_a\sin i$ & $i$ & $\alpha$\\
 & Name & & (mas) & (\degr) & (\%) &  & Name & & (mas) & (\degr) & (\%)\\ \hline
1292 & GJ 3021 & 1 & 9.86e-02 & $12.3\pm13.6$ & 65 & 64426 & 114762 & 16 & 1.18e-01 & $4.3\pm3.4$ & 41\\
5054 & 6434 & 2 & 1.79e-03 & $0.2\pm0.2$ & 51 & 65721 & 70 Vir & 17 & 1.72e-01 & $13.7\pm9.5$ & 42\\
7513b & $\upsilon$ And b & 3  & 2.30e-03 & $179.7\pm0.3$ & 62 & 67275 & $\tau$ Boo & 11 & 9.28e-03 & $0.9\pm0.8$ & 55\\
7513c & $\upsilon$ And c & 3 & 9.43e-02 & $173.7\pm3.8$ & 23 & 68162 & 121504 & 2 & 6.16e-03 & $0.3\pm0.2$ & 27\\
7513d & $\upsilon$ And d & 3 & 5.76e-01 & $28.7\pm16.8$ & 20 & 72339 & 130322 & 18 & 3.83e-03 & $0.2\pm0.2$ & 61\\
8159 & 10697 & 4 & 3.48e-01 & $169.2\pm4.2$ & 6 & 74500 & 134987 & 4 & 4.41e-02 & $2.6\pm2.8$ & 55\\
9683 & 12661 & 5 & 5.55e-02 & $3.0\pm1.9$ & 36 & 78459 & $\rho$ CrB & 19 & 1.35e-02 & $179.1\pm0.5$ & 1\\
10138 & Gl 86 & 6 & 5.06e-02 & $8.5\pm11.9$ & 75 & 79248 & 14 Her & 20 & 7.75e-01 & $140.6\pm27.1$ & 25\\
12048 & 16141 & 7 & 2.08e-03 & $0.1\pm0.1$ & 46 & 89844b & 168443b & 20 & 5.95e-02 & $178.0\pm1.8$ & 50\\
12653 & HR 810 & 8 & 1.14e-01 & $7.0\pm4.4$ & 20 & 89844c & 168443c & 20 & 1.11e+00 & $48.7\pm42.4$ & 60\\
14954 & 19994 & 2 & 8.34e-02 & $4.9\pm3.7$ & 40 & 90485 & 169830 & 12 & 4.60e-02 & $2.1\pm1.1$ & 10\\
16537 & $\epsilon$ Eri & 9 & 1.06e+00 & $174.0\pm4.1$ & 44 & 93746 & 177830 & 4 & 1.86e-02 & $1.3\pm0.7$ & 19\\
26381 & 37124 & 4 & 1.80e-02 & $179.5\pm0.4$ & 55 & 96901 & 16 Cyg B & 21 & 1.18e-01 & $170.8\pm7.2$ & 41\\
27253 & 38529 & 5 & 1.62e-03 & $0.1\pm0.0$ & 3 & 97336 & 187123 & 4 & 4.09e-04 & $180.0\pm0.0$ & 17\\
31246 & 46375 & 7 & 2.95e-04 & $0.0\pm0.0$ & 46 & 98714 & 190228 & 22 & 1.40e-01 & $5.1\pm2.4$ & 9\\
33719 & 52265 & 10 & 1.66e-02 & $178.4\pm1.5$ & 59 & 99711 & 192263 & 4 & 7.62e-03 & $179.4\pm0.5$ & 73\\
43587 & 55 Cnc & 11 & 8.38e-03 & $179.5\pm0.3$ & 30 & 100970 & 195019 & 4 & 1.28e-02 & $0.3\pm0.1$ & $<1$\\
47007 & 82943 & 12 & 8.58e-02 & $7.2\pm7.3$ & 61 & 108859 & 209458 & 23 & 5.75e-04 & $0.0\pm0.0$ & 19\\
47202b & 83443b & 13 & 3.50e-04 & $0.0\pm0.0$ & 13 & 109378 & 210277 & 4 & 6.59e-02 & $175.3\pm3.6$ & 41\\
47202c & 83443c & 13 & 7.78e-04 & $0.0\pm0.0$ & 3 & 113020 & Gl 876 & 24 & 2.82e-01 & $172.2\pm7.2$ & 49\\
50786 & 89744 & 14 & 1.06e-01 & $175.4\pm2.9$ & 35 & 113357 & 51 Peg & 25 & 1.49e-03 & $0.1\pm0.2$ & 65\\
52409 & 92788 & 2 & 1.28e-01 & $4.4\pm2.8$ & 22 & 113421 & 217107 & 4 & 4.58e-03 & $179.6\pm0.4$ & 54\\
53721 & 47 UMa & 15 & 3.24e-01 & $45.4\pm72.0$ & 86 & 116906 & 222582 & 4 & 1.58e-01 & $5.4\pm2.9$ & 10\\
\hline
\end{tabular}
\end{table*}

Fitting the IAD of HD~209458 (HIP 108859) using the spectroscopic orbit by \citet{Mazeh-2000:a} yields $i\approx 0.02$\degr\  even if we know that transits do occur \citep{Charbonneau-2000:a}, implying $i\sim 90$\degr.  We need a way to rule out such solutions.

Table \ref{Tab:Planets} lists extrasolar planet candidates with their $a_a\sin i$ based on their Hipparcos parallax and their most recent spectroscopic orbit.  For all but seven, the value of $a_a\sin i$ constrains the inclination to be $<10$\degr\ (or $>170$\degr) if one assumes that Hipparcos noticed the orbital motion.  That latter assumption is where the weakness of the reasoning lies.  An F-test can be performed to see
whether the two additional parameters improve the fit of the IAD with
respect to the single star model.  In the absence of an astrometric wobble, the quantity
\begin{equation}
\hat{F}=\frac{N-7}{2}\frac{\chi^2_S-\chi^2_C}{\chi^2_C}
\end{equation}
follows the F-distribution with $(2,N-7)$ degrees of freedom \citep{DaReErAnPhSc}.  $N$ denotes 
the number of data points and $\chi^2_S$ and $\chi^2_C$ are the value of the 
$\chi^2$ with the 5-parameter (single star) model and orbital model
respectively.  

The column labeled $\alpha$ in Table \ref{Tab:Planets} gives the probability 
of obtaining an $F$-value greater or equal to $\hat{F}$ if the null hypothesis H0: `no orbital wobble
present in the IAD' holds.  That hypothesis is rejected for all but four stars 
at a 5\% level.  Even if two rejections are expected by chance in this sample,
HD~38529 (HIP 27253), HD~83443 (HIP 47202), $\rho$ CrB (HIP 78459), and HD~195019 (HIP 100970) do deserve some further investigations.  However, a small $\alpha$ only means that an orbital model improves 
the fit of the IAD, with no guaranty that the adopted parameters indeed yield 
the best possible fit.  So the obtained inclination might be unreliable even 
when $\alpha\approx 0$.

The F-test rejects the astrometric orbital solutions for $\upsilon$ And 
(HIP~7513) and for HD~10697 as well as the face-on solution we would obtain 
for HD~209458.  It is worth pointing out that the value of $\alpha$ depends 
on the adopted parameters.  Therefore, an alternative orbit might still 
substantially improve the fit even if the value of $\alpha$ listed in 
Table \ref{Tab:Planets} is high.

One should also mention that $a_a\sin i$ {\em large} with respect to the precision of the instrument is not a sufficient condition for a reliable astrometric solution.  For instance, \citet{Pourbaix-2000:b} fitted the IAD of 81 single-lined spectroscopic binaries and obtained reliable results for only 24 of them.  For 21 out of these 24, $a_a\sin i$ ranges from 0.6 to 13.8 mas whereas only 20 among the 57 others are characterized by a $a_a\sin i$ much smaller than 1 mas.  However, 27\% of the stars rejected although they fulfill the 1 mas criterion have periods exceeding 11 years.  So no orbital solution could be derived for them due to the poor orbit coverage during the Hipparcos mission.

In the present sample, HD~168443c (HIP 89844) and $\epsilon$ Eri (HIP 16537) both have $a_a\sin i>1$~mas but fail to get a significant astrometric solution probably because of their relatively long periods (4.57 and 6.86 years, respectively).

\section{Conclusion}
Fitting $(i, \Omega)$ to the IAD when the spectroscopically constrained $a_a\sin i$ is much smaller than the astrometric precision always yields low values of $\sin i$, irrespective of the true inclination.  Although one cannot rule out the possibility of almost face-on orbits, very few of these orbits result in a significant improvement of the astrometric fit with respect to the single star model.
 
Hipparcos was a very successful mission and its files certainly still hold some hidden results.  However, if one does not take care, one may be tempted to make these observations tell more than what they actually can.  Moreover, the fact that one derives the same result with two instruments belonging to the same class of precision, for instance MAP and Hipparcos, does not always suffice to assess its reliability.  Before a 100$\mu$as-class instrument becomes available, the astrometric techniques will not derive the mass of most of the extrasolar sub-stellar companions known today.  The four stars with a formally significant astrometric orbit ($\alpha<5$\% in Table \ref{Tab:Planets}) do on the other hand deserve serious further consideration.  

\begin{acknowledgements}
I thank Lennart Lindegren, the referee, and Fr\'e\-d\'eric Arenou for their very useful comments, especially the former for pointing out an error in the original computation of $\alpha$. I also thank the National Aeronautics and Space Administration which partially supported this work via grant NAG5-6734.  This research has made use of the Simbad database operated at CDS, Strasbourg, France.
\end{acknowledgements}

\bibliographystyle{apj}

\begin{thebibliography}{39}
\expandafter\ifx\csname natexlab\endcsname\relax\def\natexlab#1{#1}\fi

\bibitem[{{Bevington} \& {Robinson}(1992)}]{DaReErAnPhSc}
{Bevington}, P.~R. \& {Robinson}, D.~K. 1992, Data reduction and error analysis
  for the physical sciences, 2nd edn. (McGraw-Hill)

\bibitem[{{Binnendijk}(1960)}]{PrDoSt}
{Binnendijk}, L. 1960, Properties of Double Stars (University of Pennsylvania
  Press)

\bibitem[{{Butler} \& {Marcy}(1996)}]{Butler-1996:a}
{Butler}, R.~P. \& {Marcy}, G.~W. 1996, ApJ, 464, L153

\bibitem[{{Butler} {et~al.}(1999){Butler}, {Marcy}, {Fischer}, {Brown},
  {Contos}, {Korzennik}, {Nisenson}, \& {Noyes}}]{Butler-1999:a}
{Butler}, R.~P., {Marcy}, G.~W., {Fischer}, D.~A., {Brown}, T.~W., {Contos},
  A.~R., {Korzennik}, S.~G., {Nisenson}, P., \& {Noyes}, R.~W. 1999, ApJ, 526,
  916

\bibitem[{{Butler} {et~al.}(1997){Butler}, {Marcy}, {Williams}, {Hauser}, \&
  {Shirts}}]{Butler-1997:a}
{Butler}, R.~P., {Marcy}, G.~W., {Williams}, E., {Hauser}, H., \& {Shirts}, P.
  1997, ApJ, 474, L115

\bibitem[{{Butler} {et~al.}(2000){Butler}, {Vogt}, {Marcy}, {Fischer}, {Henry},
  \& {Apps}}]{Butler-2000:a}
{Butler}, R.~P., {Vogt}, S.~S., {Marcy}, G.~W., {Fischer}, D.~A., {Henry},
  G.~W., \& {Apps}, K. 2000, ApJ, 545, 504

\bibitem[{{Charbonneau} {et~al.}(2000){Charbonneau}, {Brown}, {Latham}, \&
  {Mayor}}]{Charbonneau-2000:a}
{Charbonneau}, D., {Brown}, T.~M., {Latham}, D.~W., \& {Mayor}, M. 2000, ApJ,
  529, L45

\bibitem[{{Cochran} {et~al.}(1997){Cochran}, {Hatzes}, {Butler}, \&
  {Marcy}}]{Cochran-1997:a}
{Cochran}, W.~D., {Hatzes}, A.~P., {Butler}, R.~P., \& {Marcy}, G.~W. 1997,
  ApJ, 483, 457

\bibitem[{{Delfosse} {et~al.}(1998){Delfosse}, {Forveille}, {Mayor}, {Perrier},
  {Naef}, \& {Queloz}}]{Delfosse-1998:a}
{Delfosse}, X., {Forveille}, T., {Mayor}, M., {Perrier}, C., {Naef}, D., \&
  {Queloz}, D. 1998, A\&A, 338, L67

\bibitem[{{ESA}(1997)}]{Hipparcos}
{ESA}. 1997, The Hipparcos and Tycho Catalogues (ESA SP-1200)

\bibitem[{{Fischer} {et~al.}(2001){Fischer}, {Marcy}, {Butler}, {Vogt},
  {Frink}, \& {Apps}}]{Fischer-2001:a}
{Fischer}, D.~A., {Marcy}, G.~W., {Butler}, R.~P., {Vogt}, S.~S., {Frink}, S.,
  \& {Apps}, K. 2001, ApJ, (accepted)

\bibitem[{{Gatewood} {et~al.}(2000){Gatewood}, {Han}, \&
  {Black}}]{Gatewood-2001:a}
{Gatewood}, G., {Han}, I., \& {Black}, D. 2000, ApJ, 548, L61

\bibitem[{{Gatewood}(1987)}]{Gatewood-1987:a}
{Gatewood}, G.~D. 1987, AJ, 94, 213

\bibitem[{{Halbwachs} {et~al.}(2000){Halbwachs}, {Arenou}, {Mayor}, {Udry}, \&
  {Queloz}}]{Halbwachs-2000:a}
{Halbwachs}, J.~L., {Arenou}, F., {Mayor}, M., {Udry}, S., \& {Queloz}, D.
  2000, A\&A, 355, 581

\bibitem[{{Han} {et~al.}(2001){Han}, {Black}, \& {Gatewood}}]{Han-2001:a}
{Han}, I., {Black}, D.~C., \& {Gatewood}, G. 2001, ApJ, 548, L57

\bibitem[{{Hatzes} {et~al.}(2000){Hatzes}, {Cochran}, {McArthur}, {Baliunas},
  {Walker}, {Campbell}, {Irwin}, {Yang}, {K\"urster}, {Endl}, {Els}, {Butler},
  \& {Marcy}}]{Hatzes-2000:a}
{Hatzes}, A.~P., {Cochran}, W.~D., {McArthur}, B., {Baliunas}, S.~L., {Walker},
  G. A.~H., {Campbell}, B., {Irwin}, A.~W., {Yang}, S., {K\"urster}, M.,
  {Endl}, M., {Els}, S., {Butler}, R.~P., \& {Marcy}, G.~W. 2000, ApJ, 544,
  L145

\bibitem[{{Korzennik} {et~al.}(2000){Korzennik}, {Brown}, {Fischer},
  {Nisenson}, \& {Noyes}}]{Korzennik-2000:a}
{Korzennik}, S.~G., {Brown}, T.~M., {Fischer}, D.~A., {Nisenson}, P., \&
  {Noyes}, R.~W. 2000, ApJ, 533, L147

\bibitem[{{K{\"u}rster} {et~al.}(2000){K{\"u}rster}, {Endl}, {Els}, {Hatzes},
  {Cochran}, {D{\"o}bereiner}, \& {Dennerl}}]{Kurster-2000:a}
{K{\"u}rster}, M., {Endl}, M., {Els}, S., {Hatzes}, A.~P., {Cochran}, W.~D.,
  {D{\"o}bereiner}, S., \& {Dennerl}, K. 2000, A\&A, 353, L33

\bibitem[{{Marcy} \& {Butler}(1996)}]{Marcy-1996:a}
{Marcy}, G.~W. \& {Butler}, R.~P. 1996, ApJ, 464, L147

\bibitem[{{Marcy} {et~al.}(2000){Marcy}, {Butler}, \& {Vogt}}]{Marcy-2000:a}
{Marcy}, G.~W., {Butler}, R.~P., \& {Vogt}, S.~S. 2000, ApJ, 536, L43

\bibitem[{{Marcy} {et~al.}(1999){Marcy}, {Butler}, {Vogt}, {Fischer}, \&
  {Liu}}]{Marcy-1999:a}
{Marcy}, G.~W., {Butler}, R.~P., {Vogt}, S.~S., {Fischer}, D., \& {Liu}, M.~C.
  1999, ApJ, 520, 239

\bibitem[{{Mayor} {et~al.}(2000){Mayor}, {Naef}, {Queloz}, {Santos}, {Udry}, \&
  {Burnet}}]{Mayor-2000:a}
{Mayor}, M., {Naef}, D., {Queloz}, D., {Santos}, N.~C., {Udry}, S., \&
  {Burnet}, M. 2000, in Planetary Systems in the Universe: Observation,
  Formation and Evolution IAU Symposium 202 ASP Conference Series \#, ed. A.~J.
  {Penny}, P.~{Artymowicz}, A.~M. {Lagrange}, \& S.~S. {Russell}

\bibitem[{{Mayor} \& {Queloz}(1995)}]{Mayor-1995:a}
{Mayor}, M. \& {Queloz}, D. 1995, Nat, 378, 355

\bibitem[{{Mazeh} {et~al.}(2000){Mazeh}, {Naef}, {Torres}, {Latham}, {Mayor},
  {Beuzit}, {Brown}, {Buchhave}, {Burnet}, {Carney}, {Charbonneau}, {Drukier},
  {Laird}, {Pepe}, {Perrier}, {Queloz}, {Santos}, {Sivan}, {Udry}, \&
  {Zucker}}]{Mazeh-2000:a}
{Mazeh}, T., {Naef}, D., {Torres}, G., {Latham}, D.~W., {Mayor}, M., {Beuzit},
  J., {Brown}, T.~M., {Buchhave}, L., {Burnet}, M., {Carney}, B.~W.,
  {Charbonneau}, D., {Drukier}, G.~A., {Laird}, J.~B., {Pepe}, F., {Perrier},
  C., {Queloz}, D., {Santos}, N.~C., {Sivan}, J., {Udry}, S., \& {Zucker}, S.
  2000, ApJ, 532, L55

\bibitem[{{Mazeh} {et~al.}(1999){Mazeh}, {Zucker}, {Dalla {T}orre}, \& {van
  {L}eeuwen}}]{Mazeh-1999:a}
{Mazeh}, T., {Zucker}, S., {Dalla {T}orre}, A., \& {van {L}eeuwen}, F. 1999,
  ApJ, 522, L149

\bibitem[{{Naef} {et~al.}(2000){Naef}, {Mayor}, {Pepe}, {Queloz}, {Udry}, \&
  {Burnet}}]{Naef-2000:a}
{Naef}, D., {Mayor}, M., {Pepe}, F., {Queloz}, D., {Udry}, S., \& {Burnet}, M.
  2000, in Disks, Planetesimals and Planets ASP Conference Series \#, ed.
  F.~{Garz\'on}, C.~{Eiroa}, D.~{de {W}inter}, \& T.~J. {Mahoney}

\bibitem[{{Naef} {et~al.}(2001){Naef}, {Mayor}, {Pepe}, {Queloz}, {Udry}, \&
  {Burnet}}]{Naef-2001:a}
{Naef}, D., {Mayor}, M., {Pepe}, F., {Queloz}, D., {Udry}, S., \& {Burnet}, M.
  2001, A\&A, (submitted)

\bibitem[{{Noyes} {et~al.}(1999){Noyes}, {Contos}, {Korzennik}, {Nisenson},
  {Brown}, \& {Horner}}]{Noyes-1999:a}
{Noyes}, R.~W., {Contos}, A.~R., {Korzennik}, S.~G., {Nisenson}, P., {Brown},
  T.~M., \& {Horner}, S.~D. 1999, in Precise stellar radial velocities IAU
  Colloquium 170 ASP Conference Series \#185, ed. J.~B. {Hearnshaw} \& C.~D.
  {Scarfe}

\bibitem[{{Penny} {et~al.}(2000){Penny}, {Artymowicz}, {Lagrange}, \&
  {Russell}}]{IAUS202}
{Penny}, A.~J., {Artymowicz}, P., {Lagrange}, A.~M., \& {Russell}, S.~S., eds.
  2000, Planetary Systems in the Universe: Observation, Formation and Evolution

\bibitem[{{Perryman} {et~al.}(1997){Perryman}, {Lindegren}, {Arenou},
  {Bastian}, {Bernstein}, {van {L}eeuwen}, {Schrijver}, {Bernacca}, {Evans},
  {Falin}, {Froeschle}, {Grenon}, {Hering}, {H{\o}g}, {Kovalevsky}, {Mignard},
  {Murray}, {Penston}, {Petersen}, {Le {P}oole}, {S\"oderhjelm}, \&
  {Turon}}]{Perryman-1996:a}
{Perryman}, M. A.~C., {Lindegren}, L., {Arenou}, F., {Bastian}, U.,
  {Bernstein}, H.~H., {van {L}eeuwen}, F., {Schrijver}, H., {Bernacca}, P.~L.,
  {Evans}, D.~W., {Falin}, J.~L., {Froeschle}, M., {Grenon}, M., {Hering}, R.,
  {H{\o}g}, E., {Kovalevsky}, J., {Mignard}, F., {Murray}, C.~A., {Penston},
  M.~J., {Petersen}, C.~S., {Le {P}oole}, R.~S., {S\"oderhjelm}, S., \&
  {Turon}, C. 1997, A\&A, 323, L49

\bibitem[{{Pourbaix} \& {Jorissen}(2000)}]{Pourbaix-2000:b}
{Pourbaix}, D. \& {Jorissen}, A. 2000, A\&AS, 145, 161

\bibitem[{{Queloz} {et~al.}(2000{\natexlab{a}}){Queloz}, {Mayor}, {Naef},
  {Pepe}, {Santos}, {Udry}, \& {Burnet}}]{Queloz-2000:a}
{Queloz}, D., {Mayor}, M., {Naef}, D., {Pepe}, F., {Santos}, N.~C., {Udry}, S.,
  \& {Burnet}, M. 2000{\natexlab{a}}, in Planetary Systems in the Universe:
  Observation, Formation and Evolution IAU Symposium 202 ASP Conference Series
  \#, ed. A.~J. {Penny}, P.~{Artymowicz}, A.~M. {Lagrange}, \& S.~S. {Russell}

\bibitem[{{Queloz} {et~al.}(2000{\natexlab{b}}){Queloz}, {Mayor}, {Weber},
  {Bl\'echa}, {Burnet}, {Confino}, {Naef}, {Pepe}, {Santos}, \&
  {Udry}}]{Queloz-2000:b}
{Queloz}, D., {Mayor}, M., {Weber}, L., {Bl\'echa}, A., {Burnet}, M.,
  {Confino}, B., {Naef}, D., {Pepe}, F., {Santos}, N., \& {Udry}, S.
  2000{\natexlab{b}}, A\&A, 354, 99

\bibitem[{{Sivan} {et~al.}(2000){Sivan}, {Mayor}, {Naef}, {Queloz}, {Udry},
  {Perrrier-{B}ellet}, \& {Beuzit}}]{Sivan-2000:a}
{Sivan}, J.~P., {Mayor}, M., {Naef}, D., {Queloz}, D., {Udry}, S.,
  {Perrrier-{B}ellet}, C., \& {Beuzit}, J.~L. 2000, in Planetary Systems in the
  Universe: Observation, Formation and Evolution IAU Symposium 202 ASP
  Conference Series \#, ed. A.~J. {Penny}, P.~{Artymowicz}, A.~M. {Lagrange},
  \& S.~S. {Russell}

\bibitem[{{S\"oderhjelm}(1999)}]{Soderhjelm-1999}
{S\"oderhjelm}, S. 1999, A\&A, 341, 121

\bibitem[{{Udry} {et~al.}(2000{\natexlab{a}}){Udry}, {Mayor}, {Naef}, {Pepe},
  {Queloz}, {Santos}, {Burnet}, {Confino}, \& {Melo}}]{Udry-2000:b}
{Udry}, S., {Mayor}, M., {Naef}, D., {Pepe}, F., {Queloz}, D., {Santos}, N.~C.,
  {Burnet}, M., {Confino}, B., \& {Melo}, C. 2000{\natexlab{a}}, A\&A, 356, 590

\bibitem[{{Udry} {et~al.}(2000{\natexlab{b}}){Udry}, {Mayor}, \&
  {Queloz}}]{Udry-2000:a}
{Udry}, S., {Mayor}, M., \& {Queloz}, D. 2000{\natexlab{b}}, in Planetary
  Systems in the Universe: Observation, Formation and Evolution IAU Symposium
  202 ASP Conference Series \#, ed. A.~J. {Penny}, P.~{Artymowicz}, A.~M.
  {Lagrange}, \& S.~S. {Russell}

\bibitem[{{Vogt} {et~al.}(2000){Vogt}, {Marcy}, {Butler}, \&
  {Apps}}]{Vogt-2000:a}
{Vogt}, S.~S., {Marcy}, G.~W., {Butler}, R.~P., \& {Apps}, K. 2000, ApJ, 536,
  902

\bibitem[{{Zucker} \& {Mazeh}(2000)}]{Zucker-2000:a}
{Zucker}, S. \& {Mazeh}, T. 2000, ApJ, 531, L67

\end{thebibliography}

\end{document}